\documentclass[a4paper]{jpconf}
\usepackage{graphicx}
\IfFileExists{srcltx.sty}{\usepackage[active]{srcltx}}%

\begin{document}

%%%%%%%%%%%%%%%%%%%%%%%%%%%%%%%%%

% 6 p.

\title{Sterile neutrino states}

\author{Alexander Kusenko}

\address{Department of Physics and Astronomy,
University of California, Los Angeles, CA 90095, USA
}

% \ead{}

\begin{abstract}

Neutrino masses are likely to be a manifestation of the right-handed, or
sterile neutrinos.  The number of sterile neutrinos and the scales of their
Majorana masses are unknown.  We explore theoretical arguments in favor of
the high and low scale seesaw mechanisms, review the existing experimental
results, and discuss the astrophysical hints regarding sterile neutrinos.  

\end{abstract}

\section{Sterile neutrinos in particle physics}

The term {\em sterile neutrino} was coined by Bruno~Pontecorvo, who
hypothesized the existence of the right-handed neutrinos in a seminal
paper~\cite{pontecorvo}, in which he also considered vacuum neutrino
oscillations in the laboratory and in astrophysics, the lepton number
violation, the neutrinoless double beta decay, some rare processes, such as
$\mu \rightarrow e \gamma$, and several other questions that have dominated the
neutrino physics for the next four decades.   Most models of the neutrino
masses introduce sterile (or right-handed) neutrinos to generate the masses of
the ordinary neutrinos via the seesaw mechanism~\cite{seesaw}.  The seesaw
lagrangian 
\begin{equation}
{\cal L}
  = {\cal L_{\rm SM}}+\bar N_{a} \left(i \gamma^\mu \partial_\mu 
\right )
N_{a}
  - y_{\alpha a} H \,  \bar L_\alpha N_{a} 
  - \frac{M_{a}}{2} \; \bar {N}_{a}^c N_{a} + h.c. \,,
\label{lagrangianM}
\end{equation}  
where $ {\cal L_{\rm SM}}$ is the lagrangian of the Standard Model, includes
some number $n$  of singlet neutrinos $N_a$ with Yukawa couplings $ y_{\alpha
a}$.  Here $H$ is the Higgs doublet and $L_\alpha$
($\alpha=e,\mu,\tau$) are the lepton doublets. Theoretical considerations do
not constrain the number $n$ of sterile neutrinos. In particular, there is no
constraint based on the anomaly cancellation because the sterile fermions do
not couple to the gauge fields. The experimental limits exist only for the
larger mixing angles~\cite{sterile_constraints}. To explain the neutrino masses
inferred from the atmospheric and solar neutrino experiments, $n=2$ singlets
are sufficient~\cite{2right-handed}, but a greater number is required if the
lagrangian (\ref{lagrangianM}) is to explain the LSND~\cite{deGouvea:2005er},
the r-process nucleosynthesis~\cite{r}, the pulsar 
kicks~\cite{ks97,Kusenko:review}, dark
matter~\cite{dw,Fuller,shi_fuller,nuMSM}, and the formation of supermassive
black holes~\cite{biermann_munyaneza}. 

The scale of the right-handed Majorana masses $M_{a}$ is
unknown; it can be much greater than the electroweak scale~\cite{seesaw}, 
or it may be as low as a few eV~\cite{deGouvea:2005er,deGouvea:2006gz}.   The
seesaw mechanism~\cite{seesaw} can explain the smallness of the neutrino masses
in the presence of the Yukawa couplings of order one if the
Majorana masses $M_a$ are much larger than the electroweak scale. Indeed, in
this case the masses of the lightest neutrinos are suppressed by the ratios $
\langle H \rangle/M_a$.  

However, the origin of the Yukawa couplings remains unknown, and there is no
experimental evidence to suggest that these couplings must be of order 1. In
fact, the Yukawa couplings of the charged leptons are much smaller than 1. For
example, the Yukawa coupling of the electron is as small as $10^{-6}$.  

One can ask whether some theoretical models are more likely to produce the
numbers of order one or much smaller than one.  The two possibilities are, in
fact, realized in two types of theoretical models.  If the Yukawa couplings
arise as some topological intersection numbers in string theory, they are
generally expected to be of order one~\cite{Candelas:1987rx}, although very
small couplings are also possible~\cite{Eyton-Williams:2005bg}. If the Yukawa
couplings arise from the overlap of the wavefunctions of fermions located on
different branes in extra dimensions, they can be exponentially suppressed and
are expected to be very small~\cite{Mirabelli:1999ks}.  

In the absence of the fundamental theory, one may hope to gain some insight 
about the size of the Yukawa couplings using 't~Hooft's naturalness
criterion~\cite{tHooft}, which states essentially that a number can be
naturally small if setting it to zero increases the symmetry of the lagrangian.
 A small breaking of the symmetry is then associated with the small non-zero
value of the parameter.  This naturalness criterion has been applied to a
variety of theories; it is, for example, one of the main arguments in favor of
supersymmetry. (Setting the Higgs mass to zero does not increase the symmetry
of the Standard Model.  Supersymmetry relates the Higgs mass to the Higgsino
mass, which is protected by the chiral symmetry.  Therefore, the light Higgs
boson, which is not natural in the Standard Model, becomes natural in theories
with softly broken supersymmetry.) In view of 't~Hooft's criterion, the
\textit{small} Majorana mass is natural because setting $M_a$ to zero increases
the symmetry of the lagrangian
(\ref{lagrangianM})~\cite{Fujikawa:2004jy,deGouvea:2005er}.  

One can ask whether cosmology can provide any clues as to whether the mass
scale of sterile neutrinos should be above or below the electroweak scale.  It
is desirable to have a theory that could generate the matter--antimatter
asymmetry of the universe. In both limits of large and small $M_a$ one can have
a successful leptogenesis: in the case of the high-scale seesaw, the baryon
asymmetry can be generated from the out-of-equilibrium decays of heavy
neutrinos~\cite{Fukugita:1986hr}, while in the case of the low-energy seesaw,
the matter-antimatter asymmetry can be produced by  
the neutrino oscillations~\cite{baryogenesis}.  The
Big-Bang nucleosynthesis (BBN) can provide a constraint on the number of light
relativistic species in equilibrium~\cite{bbn}, but the sterile neutrinos with
the small mixing angles may never be in equilibrium in the early universe, even
at the highest temperatures~\cite{dw}.  Indeed, the effective mixing angle of
neutrinos at high temperature is suppressed due to the interactions with
plasma~\cite{high-T}, and, therefore, the sterile neutrinos may never
thermalize.  High-precision measurements of the primordial abundances may probe
the existence of sterile neutrinos and the lepton asymmetry of the universe in
the future~\cite{Smith:2006uw}.  

While many seesaw models assume that the sterile neutrinos have very large
masses, which makes them unobservable, it is worthwhile to consider light
sterile neutrinos in view of the above arguments, and also because they can
explain several experimental results.  In particular, sterile neutrinos can
account for cosmological dark matter~\cite{dw}, they can explain the observed
velocities of pulsars~\cite{ks97,Kusenko:review}, the x-ray photons from
their decays can affect the star formation~\cite{reion}.  Finally, sterile
neutrinos can explain the LSND
result~\cite{deGouvea:2005er,Sorel:2003hf,LSND_sterile_decay},
which is currently being tested by the MiniBooNE experiment.

\section{Experimental limits}

Laboratory experiments are able to set limits or discover sterile neutrinos
with a large enough mixing angle.  Depending on the mass, they
can be searched in different experiments. 

The light sterile neutrinos, with masses below $10^2$~eV, can be discovered in
one of the neutrino oscillations experiments~\cite{Smirnov:2006bu}.  In fact,
LSND has reported a result~\cite{LSND}, which, in combination with the other
experiments, implies the existence of at least one sterile neutrino, more
likely, two sterile
neutrinos~\cite{deGouvea:2005er,Sorel:2003hf}.  It is
also possible that sterile neutrino decays, rather than oscillations, are the 
explanation of the LSND result~\cite{LSND_sterile_decay}.

In the eV to MeV mass range, the ``kinks'' in the spectra of beta-decay
electrons can be used to set limits on sterile neutrinos mixed with the
electron neutrinos~\cite{Shrock81}.  Neutrinoless double beta decays can probe
the Majorana neutrino masses~\cite{Elliott:2002xe}. 

For masses in the MeV--GeV range, peak searches in production of neutrinos
provide the limits.  The massive neutrinos $\nu_i$, if they exist, can be 
produced in meson decays, e.g. $\pi^\pm \rightarrow \mu^\pm \nu_i$,  with
probabilities that depend on the mixing in the charged current.  The
energy spectrum of muons in such decays should contain monochromatic
lines~\cite{Shrock81} at
$ %\begin{equation}
T_i = ( m_\pi^2 + m_\mu^2 - 2 m_\pi m_\mu - m_{\nu_i}^2) / 2 m_\pi. 
% \label{spectrummu}
$ % \end{equation}
Also, for the MeV--GeV masses one can set a number of constraints based on the
decays of the heavy neutrinos into the ``visible'' particles, which would
be observable by various detectors. These limits are discussed in
Ref.~\cite{sterile_constraints}.

\section{Sterile neutrinos in astrophysics and cosmology}

\begin{figure}[h]
\includegraphics[width=20pc]{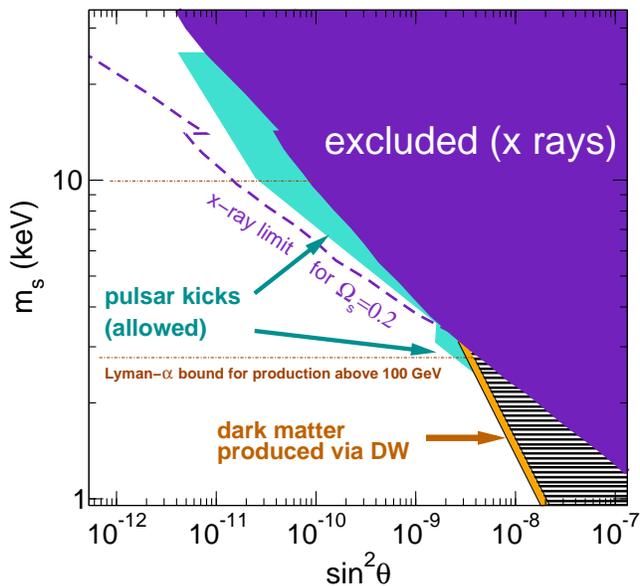}\hspace{2pc}
\begin{minipage}[b]{15pc}\caption{\label{fig:range} Sterile neutrinos in the
keV mass range can be dark matter; their emission from a supernova can
explain the observed velocities of pulsars.  If the sterile neutrinos account
for all dark matter, they must be sufficiently cold to satisfy the
cosmological bounds on the mass.  The limit depends on the
production mechanism in the early universe.  The lower
bound of 2.7~keV corresponds to production at the
electroweak scale~\cite{Kusenko:2006rh}. \ \\ \ \\ \ }
\end{minipage}
\end{figure}

Sterile neutrinos can be produced in supernova explosions.  The observations of
neutrinos from SN1987A constrain the amount of energy that the sterile
neutrinos can take out of the supernova, but they are still consistent with the
sterile neutrinos that carry away as much as a half of the total energy of the
supernova.  A more detailed analysis shows that the emission of sterile
neutrinos from a cooling newly born neutron star is anisotropic due to the
star's magnetic field~\cite{ks97}.  The anisotropy of this emission can result
in a recoil velocity of the neutron star as high as $\sim 10^3$km/s. This
mechanism can be the explanation of the observed pulsar
velocities~\cite{Kusenko:review}. The range of masses and mixing angles
required to explain the pulsar kicks is shown in Fig.~\ref{fig:range}.  

The sterile neutrinos could play an important role in Big-Bang
nucleosynthesis~\cite{Smith:2006uw}, as well as in the synthesis of heavy
elements in the supernova, by enhancing the r-process~\cite{r}.  

The sterile neutrinos can be the cosmological dark
matter~\cite{dw,Fuller,shi_fuller,nuMSM}.  The interactions already
present in the lagrangian (\ref{lagrangianM}) allow for the production of relic
sterile neutrinos via the Dodelson-Widrow (DW) mechanism~\cite{dw} in the right
amount to account for all dark matter, i.e. $\Omega_s\approx 0.2$.  The
x-ray limits on the photons from the decays of the relic sterile
neutrinos~\cite{x-rays} forces them to have mass of at least a few keV if they
are produced a la DW. However, these neutrinos appear to be too warm to agree
with the Lyman-$\alpha$ bound~\cite{viel}, which is $m_s>10$~keV in this
scenario (see Fig.~\ref{fig:range}).

If the lepton asymmetry of
the universe is relatively large, the resonant oscillations can produce the
requisite amount of dark matter even for smaller mixing angles, for which
there are no x-ray limits. (The x-ray flux is proportional to the square of
the mixing angle.)  

It is also possible that some additional interactions, not
present in eq.~(\ref{lagrangianM}) can be responsible for the production of
dark-matter sterile neutrinos~\cite{shaposhnikov_tkachev,Kusenko:2006rh}. 
For example, if the mass $M\sim$~keV is not a fundamental constant of nature,
but is the result of some symmetry breaking via the Higgs mechanism, the
Lyman-$\alpha$ bound can be relaxed to well below the current
x-ray limits~\cite{Kusenko:2006rh}.  In this case the same sterile neutrino
can simultaneously explain the pulsar kicks and dark matter
(Fig.~\ref{fig:range}).  The Higgs field giving the sterile neutrinos their
Majorana mass, the so called \textit{singlet Majoron}~\cite{singlet_Majoron},
can be discovered at the Large Hadron Collider (LHC). 

As was mentioned above, the relic sterile neutrinos can decay into the lighter
neutrinos and an the x-ray photons~\cite{pal_wolf}, which can be detected by
the x-ray telescopes~\cite{x-rays}. 
The flux of x-rays depends on the sterile neutrino abundance.  If all the dark
matter is made up of sterile neutrinos, $ \Omega_s\approx 0.2 $, then the
limit on the mass and the mixing angle is given by the dashed line in
Fig.~\ref{fig:range}. 
However, the interactions in the lagrangian (\ref{lagrangianM}) cannot produce
such an $ \Omega_s= 0.2 $ population of sterile neutrinos for the masses and
mixing angles along this dashed line, unless the universe has a relatively
large lepton asymmetry~\cite{shi_fuller}.  If the lepton asymmetry is small,
the interactions in eq.~(\ref{lagrangianM}) can produce the relic sterile only
via the neutrino oscillations off-resonance at some sub-GeV
temperature~\cite{dw}. This mechanism provides the lowest possible abundance
(except for the low-temperature cosmologies, in which the universe is never
reheated above a few MeV after inflation~\cite{low-reheat}).  The
model-independent bound~\cite{Kusenko:2006rh} based on this scenario is shown
as a solid (purple) region in Fig.~\ref{fig:range}.  It is based on the flux
limit from Ref.~\cite{x-rays} and the analytical fit to the numerical
calculation of sterile neutrino production by
Abazajian~\cite{Abazajian:2005gj}.  This calculation may have some
hadronic uncertainties~\cite{Asaka:2006rw}

Of course, the sterile neutrinos can have some additional
couplings~\cite{shaposhnikov_tkachev,Kusenko:2006rh}, and the additional
production can take place at higher temperatures.  In particular, if the relic
sterile neutrinos are produced above the electroweak scale, the Lyman-$\alpha$
bound is relaxed from 10~keV to 2.7~keV~\cite{Kusenko:2006rh}.  Of course, if
the sterile neutrinos constitute only a small part of dark matter, the
Lyman-$\alpha$ bounds do not apply at all.

 The x-ray photons from sterile neutrino decays in the early universe could
have affected the star formation.  Although these x-rays alone are not
sufficient to reionize the universe, they can catalyze
the production of molecular hydrogen and speed up the star
formation~\cite{reion}, which, in turn, could cause the reionization.
Molecular hydrogen is a very important
cooling agent necessary for the collapse of primordial gas clouds that gave
birth to the first stars.  The fraction of molecular hydrogen must exceed a
certain minimal value for the star formation to begin~\cite{Tegmark:1996yt}. 
The reaction H+H$\rightarrow$H$_2 +\gamma$ is very slow in comparison with the
combination of reactions  
${\rm H}^{+}+{\rm H}  \rightarrow  {\rm H}_2^++ \gamma$ and  
${\rm H}_2^{+}+{\rm H}  \rightarrow  {\rm H}_2+{\rm H}^+$, 
which are possible if the hydrogen is ionized.  Therefore, the ionization
fraction determines the rate of molecular hydrogen production.  If dark
matter is made up of sterile neutrinos, their decays produce a sufficient flux
of photons to increase the ionization fraction by as much as two orders of
magnitude~\cite{reion}.  This has a dramatic effect on the
production of molecular hydrogen and the subsequent star formation.

\section{Conclusions}

The underlying physics responsible for the neutrino masses is likely to involve
right-handed, or sterile neutrinos.  The Majorana masses of these states
can range from a few eV to values well above the electroweak
scale.  Theoretical arguments have been made in favor of both the high-scale
and low-scale seesaw mechanisms: the high-scale seesaw may be favored by the
connection with the Grand Unified Theories, while the low-scale seesaw is
favored by 't~Hooft's naturalness criterion.  Cosmological considerations are
consistent with a vast range of mass scales.  The laboratory bounds do not
provide significant constraints on the sterile neutrinos, unless they have a
large mixing with the active neutrinos.  The atmospheric and solar neutrino
oscillation results cannot be reconciled with the LSND result, unless sterile
neutrinos (or other new physics) exist.  

There are several indirect astrophysical hints in favor of sterile neutrinos
at the keV scale.  Such neutrinos can explain the observed velocities of
pulsars, they can be the dark matter, and they can play a role in star
formation and reionization of the universe. 

The preponderance of indirect astrophysical hints may be a precursor of a major
discovery, although it may also be a coincidence.  One can hope to discover
the sterile neutrinos in the x-ray observations. The mass around 3~keV and the
mixing angle  $ \sin^2 \theta \sim 3\times 10^{-9}$ appear to be particularly
interesting because the sterile neutrino with such parameters could
simultaneously explain the pulsar kicks and dark matter (assuming the sterile
neutrinos are produced at the electroweak scale).  However, it is worthwhile to
search for the signal from sterile dark matter in other parts of the allowed
parameter space shown in Fig.~\ref{fig:range}.  The existence of a much lighter
sterile neutrino, with a much greater mixing angle can be established
experimentally if MiniBooNE confirms the LSND result.

\ack
This work was supported in part by the DOE grant DE-FG03-91ER40662 and by the
NASA ATP grants NAG~5-10842 and NAG~5-13399.  The author thanks Theory Unit at
CERN and ITP, EPFL, for hospitality.

\end{document}